\documentclass{IEEEtran}
\newtheorem{definition}{Definition}
\newtheorem{proposition}{Proposition}

\usepackage{graphicx,graphics}
\usepackage{booktabs}
\usepackage{authblk}
\usepackage{amsmath}
\usepackage{subfigure}
\usepackage{textcomp}
\usepackage{soul}
\usepackage{multirow}
\usepackage{diagbox}
 \usepackage[table,xcdraw]{xcolor}

\usepackage[noadjust]{cite}
\begin{document}
%
\title{Load-balanced Routing for Nested Interconnection Networks}
%
%
%

\author{Zhipeng Xu, Xiaolong Huang and Yuefan Deng,~\IEEEmembership{Member,~IEEE}
\thanks{Z. Xu is with School of Data and Computer Science, Sun Yat-sen University, Guangzhou, Guangdong 510275, P.R. China and the Department of Applied Mathematics and Statistics, Stony Brook University, Stony Brook, NY 11794, USA. E-mail:xuzhp9@mail2.sysu.edu.cn}
\thanks{X. Huang is with the Department of Applied Mathematics and Statistics, Stony Brook University, Stony Brook, NY 11794, USA. E-mail: xiaolong.huang@stonybrook.edu}
\thanks{Y. Deng is with the Department of Applied Mathematics and Statistics,
Stony Brook University, Stony Brook, NY 11794, USA. E-mail: yuefan.deng@stonybrook.edu}
\thanks{Manuscript received XX XX, 2019; revised XX XX, 2020.}}

%
%

\markboth{MANUSCRIPT ID}{Zhipeng Xu \MakeLowercase{\textit{et al.}}: Load-balanced Routing for Nested Interconnection Networks}

%



\maketitle

\begin{abstract}
We introduced the load-balanced routing algorithms, for interconnection networks resulting from nesting, by considering the pressure of the data forwarding in each node. Benchmarks on a small cluster with various network topologies, and simulations for several larger clusters whose prototypes are too costly to construct, demonstrated substantial gains of communication performance with our routing on these networks over other mainstream routing algorithms.  
\end{abstract}


%
\IEEEpeerreviewmaketitle

\section{Introduction}
Supercomputer technologies advance dramatically during the past decades and the exascale systems to be delivered in 2020 is a milestone. The speed of a single processor following the Moore's law \cite{Moore1998} approaches a hard wall, caused by fundamental physics. To pursue extreme speed, supercomputers as shown in TOP 500 list \cite{top500} possess tens of millions of cores, and particularly, the number of cores for exascale supercomputers will approach hundreds of millions, challenges, among others, the walls of efficiency and scalability for interconnection networks. The hypercube \cite{Efe1991,Harary1988}, along with torus \cite{Adiga2005,Ajima2009,Kini2010}, fat-tree \cite{Leiserson1985,Liao2015}, and Dragonfly \cite{Kim2008,Teh2017}, widely implemented, provide powerful topologies for overcoming many scalability obstacles. Among these topologies, hypercube and tours are direct networks, similar to the peer-to-peer network, the messages can be exchanged directly between nodes on the network without external devices, \textit{e.g.}, servers or switchers. On the contrary, some mainstream indirect networks, such as fat-tree and the Dragonfly, do require such external devices.
\par Diameter of the network topology, a significant metric for assessing communication latency, is a critical measure to minimize for gaining communication efficiency. Finding graphs with small diameters, is a computational challenging task. The degree/diameter \cite{miller2005moore} problem aims at discovering the largest graph $G$ with largest degree $k$ and minimal diameter $D$ with any vertices. The Graph Golf \cite{golf}, an international competition, promotes search for minium-diameter graphs for every order/degree pair. Deng \textit{et al.} \cite{YUEFAN2012} and Sabino \textit{et al.} \cite{Sabino2018} used Lie algebra and symplectic algebra to design and optimize interconnection networks. The hierarchical network is another option and is becoming the mainstream constructs of most supercomputer networks. Dragonfly is a hierarchical network, dividing links into intra-group and inter-group, requiring optimization by parts, while fat-tree is a multi-level and multi-stage interconnection network \cite{Wu1980}. Hypercube, cross-cube \cite{Haqa}, folded-Heawood \cite{jan2004,El-Barr2011}, hyper-Petersen, and folded-Petersen \cite{Dasa} are a class of networks, resulting from Cartesian product \cite{Day1997,Xu2005,Youssefa} operation by combining graphs, a special class of hierarchical networks. To reduce the number of links between base graphs, a class of recursive networks including the extended hypercubes \cite{Kumar1992}, hierarchical Petersen network  \cite{Seo2017}, and recursive expand Heawood \cite{Wang2007} have been also proposed. Additionally, block shift network \cite{Pan1997}, hierarchical folded-hypercube network \cite{Shi2000}, hierarchical star \cite{Shi2005}, hierarchical hypercube network \cite{Malluhi1994}, interlaced bypass torus network \cite{Zhang2011,Feng2012} and Dandelion \cite{Fu2014}, are also novel with their properties, for applications in supercomputers and data centers.
\par Most hierarchical networks are based on the cube graph, Peterson graph, Heawood graph, and Hoffman–Singleton graph \cite{Hafner2003} of orders of 4, 10, 14 and 50. For graphs of order 16 and 32, we propose nested networks, resulting from multiplying the optimal base graphs of arbitrary orders, with low-radix and small MPL, that are scalable.
\par Routing for these interconnection networks is critical for their efficiency, even for optimal topologies. Usually, routing algorithms are deterministic, oblivious, and adaptive based on the selection of messaging path of the involving node pairs \cite{William2003}. In practice, shortest-length path routing, as a secondary consideration in routing algorithm \cite{William2003}, without channel load balancing cuts throughput. Valiant's algorithm \cite{Valiant1981} is a classical method by inserting an intermediate node randomly to balance loads for channels, although it does not guarantee minimum hops of any source-destination pair. The locality-preserving randomized oblivious \cite{Singh2002} algorithm, globally oblivious adaptive locally algorithm \cite{Singh}, and globally adaptive load-balanced routing \cite{Singh2004} help optimize forwarding load. Deterministic routing is more common for its easy of deployment and implementation, while adaptive routing, usually more efficient, introduces out-of-order packets and consumes more channel resources \cite{William2003}. Deterministic routing and adaptive routing achieve similar performance in some scenarios for fat-tree topology \cite{Leiserson1985}.
\par To evaluate our routing algorithm, we recalibrated the Graph500 \cite{Checconi2013} benchmark that stresses all-to-all communication. Our benchmarks are conducted on a small Beowulf cluster \cite{BE2003} with 32 nodes, while we adopted the SimGrid, a framework to evaluate platform configurations, algorithmic approaches, and system design, to asses several larger networks. Our benchmarks show that new routing algorithms improve network performance dramatically for applications that require heavy communications. The orchestrated impact of load-balanced routing with the nested network is the core contribution of our manuscript.
\par The rest of the manuscript is organized as follows: Sec. 2 introduces optimal regular graphs and discusses the topological properties of product graphs. Sec. 3 shows benchmarks of networks, resulting from a Beowulf cluster and SimGrid simulations. Sec. 4 concludes the research.
\section{Topology and routing}
\subsection{Optimal Regular Graphs}
A regular graph $(N,k)$ of $N$ vertices $k$ degrees, that has minimal mean path length (MPL) is defined as an optimal regular graph in our manuscript. Cerf \textit{et al.} \cite{Cerf1974} proved that the diameter of any regular graph with minimal MPL is also minimal.
\begin{table}[]
\caption{Properties of several optimal graphs with $N=10,14,...,32$, $k=3,4$.}
\label{tabg}
\begin{center}
\begin{tabular}{@{}llllll@{}}
\toprule
Graph       & $N$ & $k$ & $D$ & MPL  & BW \\ \midrule
Peterson \cite{holton1993petersen}       & 10       & 3      & 2        & 1.67 & 5 \\
Heawood \cite{Brinkmann2013}        & 14       & 3      & 3        & 2.07 & 7   \\
Levi \cite{Randi1979}           & 30       & 3      & 4        & 2.86 & 9   \\
(16,3)-optimal \cite{Deng2019} & 16       & 3      & 3        & 2.20 & 6   \\
(16,4)-optimal \cite{Deng2019} & 16       & 4      & 3        & 1.75 & 12  \\
(32,3)-optimal \cite{Deng2019} & 32       & 3      & 4        & 2.94 & 10   \\
(32,4)-optimal \cite{Deng2019} & 32       & 4      & 3        & 2.94 & 16   \\ \bottomrule
\end{tabular}
\end{center}
\end{table}
\par
\begin{figure}[htbp!]
\centering
\includegraphics[width=9cm]{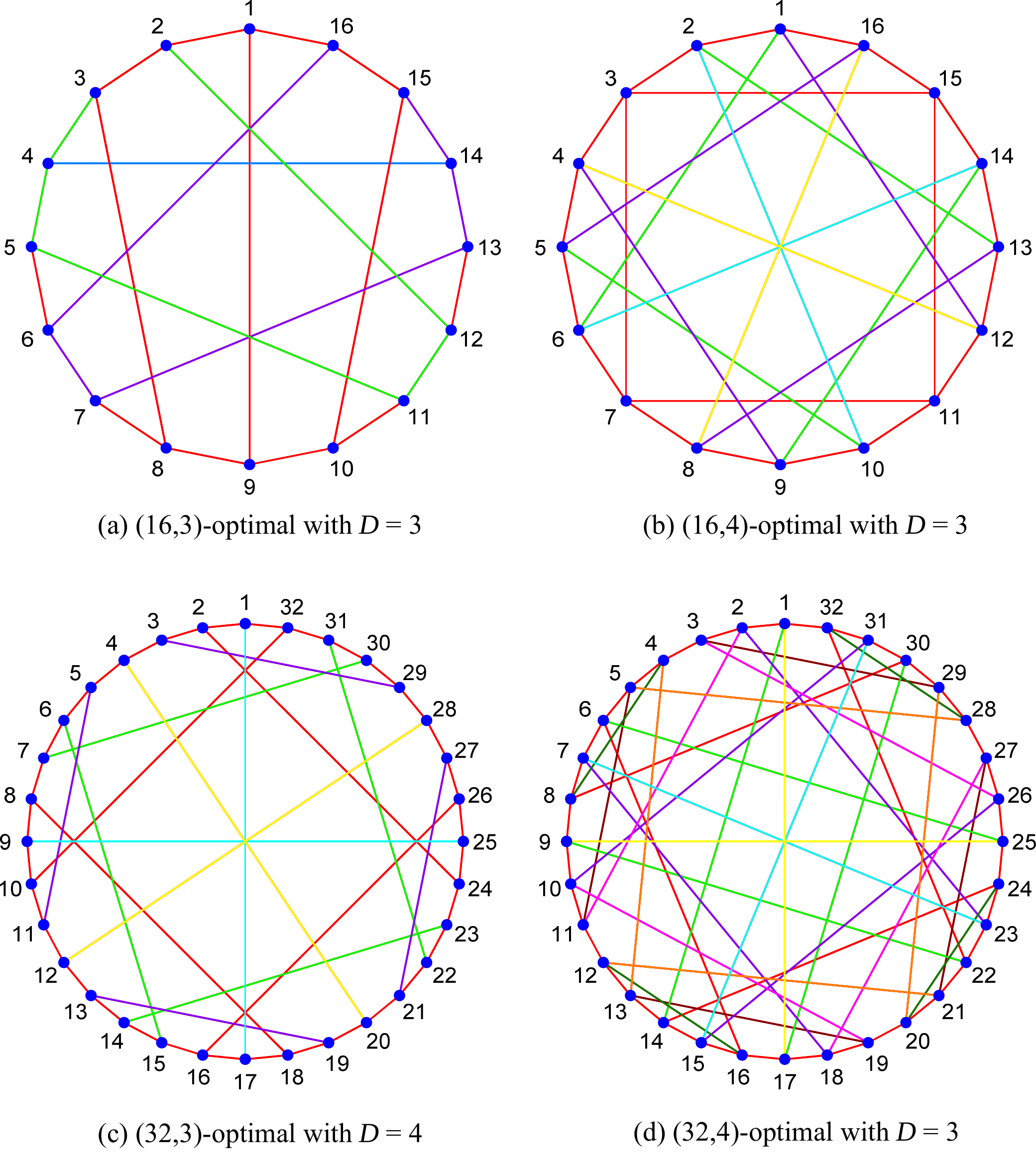}
\caption{The optimal graphs used by a group \cite{Deng2019} constructing a cluster.}\label{deng}
\end{figure}

 Brute-force search for optimal graphs with small $N$ is effective. Deng \textit{et al.} \cite{Deng2019} reported graphs with minimal MPL for $(16,3)$,$(16,4)$,$(32,3)$ and $(32,4)$ by using parallel exhaustive search \cite{Deng2019}. Figure \ref{deng} shows the optimal graphs for $N=16,32$, while Table \ref{tabg} shows $N,k,D,\mathrm{MPL}$ and bisection width (BW) of these classical low-radix graphs. 
\par When $N$ is large, the cost of search is too high and heuristic methods become the reality. Mizuno \textit{et al.} \cite{Mizuno2016} reported the result for $(256,22)$. Obviously, exhausted and heuristic methods, generating ugly graphs are still too costly. Hence, we propose the approach of the Cartesian product by nesting small optimal graphs essentially to arbitrary large and usually beautiful graphs.
\subsection{Product Networks}
\begin{figure}[htbp!]
\begin{center}
	\includegraphics[width=9cm]{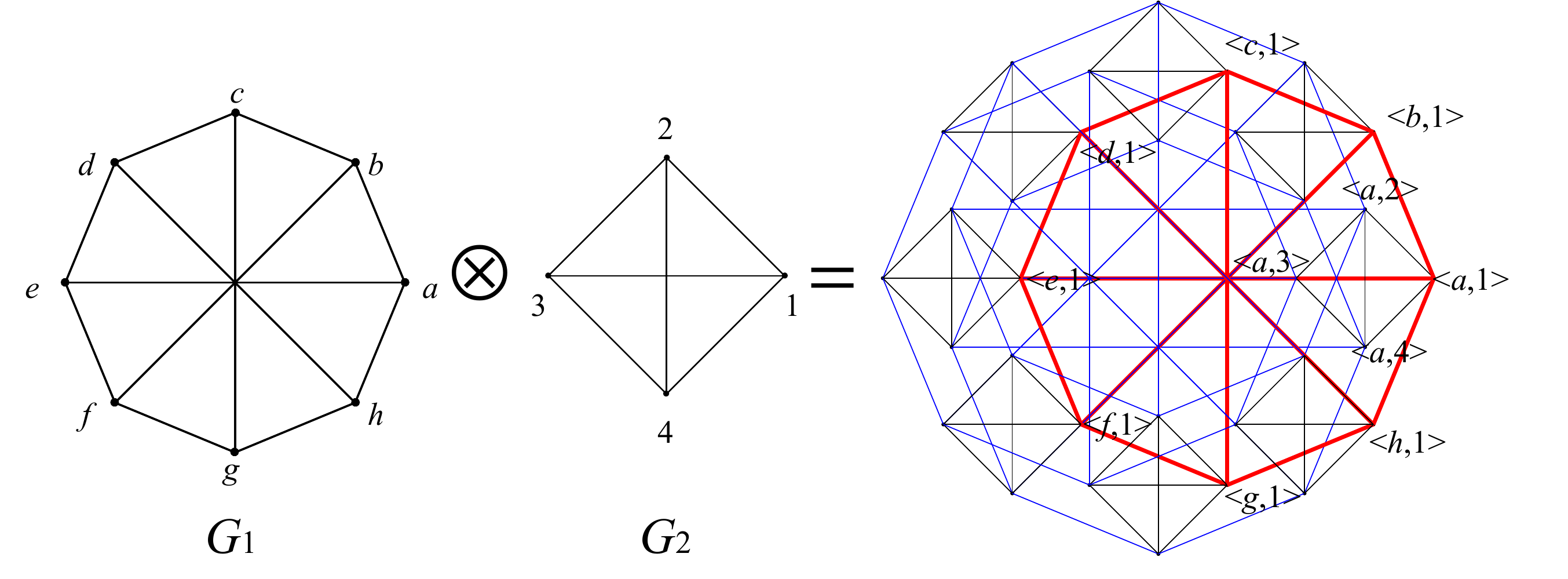}
	\caption{An example of nesting two figures $(8,3)\otimes(4,3)$.}
	\label{product}
	\end{center}
\end{figure}

\begin{definition}
The Cartesian product $G=G_1\otimes G_2$ of two undirected connected graphs $G_1=(V_1,E_1)$ and $G_2=(V_2,E_2)$ is also the undirected graph $G=(V,E)$, where the vertex for $G$ can be presented as a two-tuples $\langle u_1,u_2\rangle$ and
$V=\{\langle u_1,u_2\rangle|u_1\in V_1~\text{and}~u_2 \in V_2\}$, and the set of edges is $E=\{(\langle u_1,u_2\rangle,\langle v_1,v_2\rangle)|((u_1=v_1)~\text{and}~(u_2,v_2)\in E_2)~\text{or}~((u_2=v_2)~\text{and}~(u_1,v_1)\in E_1)\}$.
\end{definition}
The Cartesian product method is fast to construct a larger scale interconnection network as specified \cite{Xu2002}. Figure \ref{product} shows an example of the Cartesian product for $(8,3)$ and 4-vertex complete graph, where all edges for $\langle *,1\rangle$ labelled as bold red lines are organized with $G_1$, then repeated in blue lines. In addition, all $\langle a,*\rangle$ are all connected same with $G_2$. For the folded graph where $G_1=G_2$, such as the folded Peterson Graph($FP_n$) \cite{Ohringa}, we observe particular properties.
\par
The Cartesian product posses desirable properties:
\begin{itemize}
	\item the size $S(G_1\otimes G_2)=S(G_1)\cdot S(G_2)$,
    \item the degree $d(G_1\otimes G_2)=d(G_1)+d(G_2)$,
    \item the diameter $\delta(G_1\otimes G_2)=\delta(G_1)+\delta(G_2)$,
    \item $G_1\otimes G_2$ is isomorphic to $G_2\otimes G_1$.
\end{itemize}
\par
These properties, with elegant proofs \cite{Day1997,Al-Ayyoub2002}, imply that the diameter grow linearly while the scale of the network increases exponentially, and they fit well with the requirements of large scale interconnection networks. 
\par
\begin{figure}[htbp!]
\begin{center}
	\includegraphics[width=8cm]{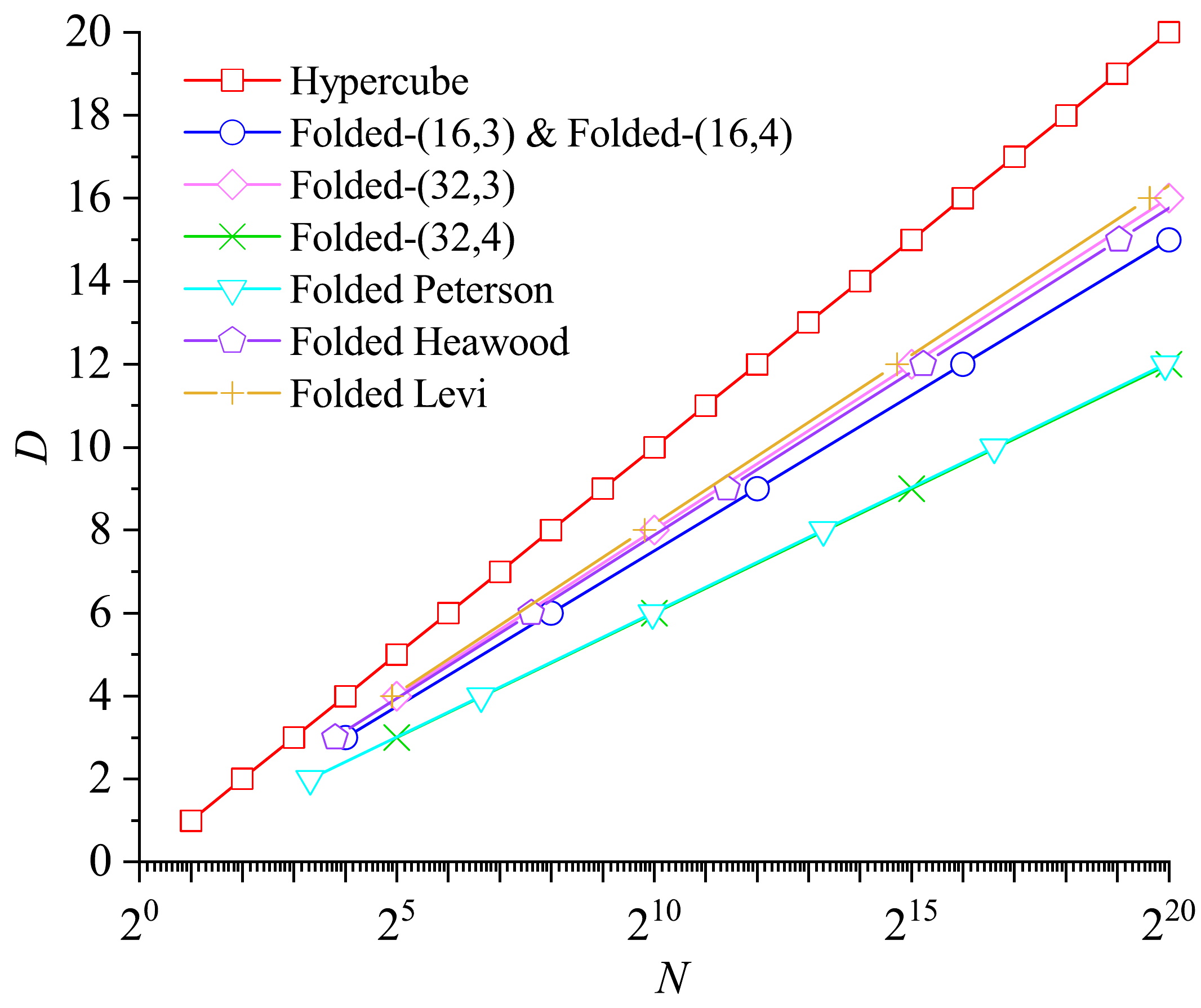}
	\caption{The diameters of six folded networks and a hypercube.}
	\label{diamter}
	\end{center}
\end{figure}

Peforming Cartesian product with optimal graphs \cite{Deng2019} that we discovered and verified, we can build scalable networks. In Fig. \ref{diamter} and Tab. \ref{diatable}, nested networks such as hypercube provide the most diversities of scales, and folded Peterson, folded-(32,4) networks that have minimal diameters, and other networks balance the diameters and diversities. The folded-(16,4) and folded-(16,3) having equal diameter, but perform differently because folded-(16,3) congest more frequently due to fewer ports.
\begin{table}[]
	\caption{The basic properties of several folded networks.}\label{diatable}
	\begin{center}
	\begin{tabular}{@{}llll@{}}
		\toprule
		Topology        & $N$ & $D$ & $k$   \\ \midrule
		Hypercube       & $2^\alpha$    & $\alpha$      & $\alpha$     \\
		Folded Peterson \cite{Ohring1993} & $10^\alpha$   & $2\alpha$     & $3\alpha$    \\ 
		Folded Heawood \cite{jan2004}       & $14^\alpha$    & $3\alpha$      & $3\alpha$     \\
		Folded Levi      & $30^\alpha$    & $4\alpha$      & $3\alpha$      \\
		Folded-(16,3)    & $16^\alpha$ & $3\alpha$ & $3\alpha$ \\
		Folded-(16,4)    & $16^\alpha$ & $3\alpha$ & $4\alpha$ \\
		Folded-(32,3)    & $32^\alpha$ & $4\alpha$ & $3\alpha$ \\
		Folded-(32,4)    & $32^\alpha$ & $3\alpha$ & $4\alpha$ \\
		\bottomrule
	\end{tabular}
\end{center}
\end{table}
\par
\begin{figure}[htbp!]
\begin{center}
	\includegraphics[width=8cm]{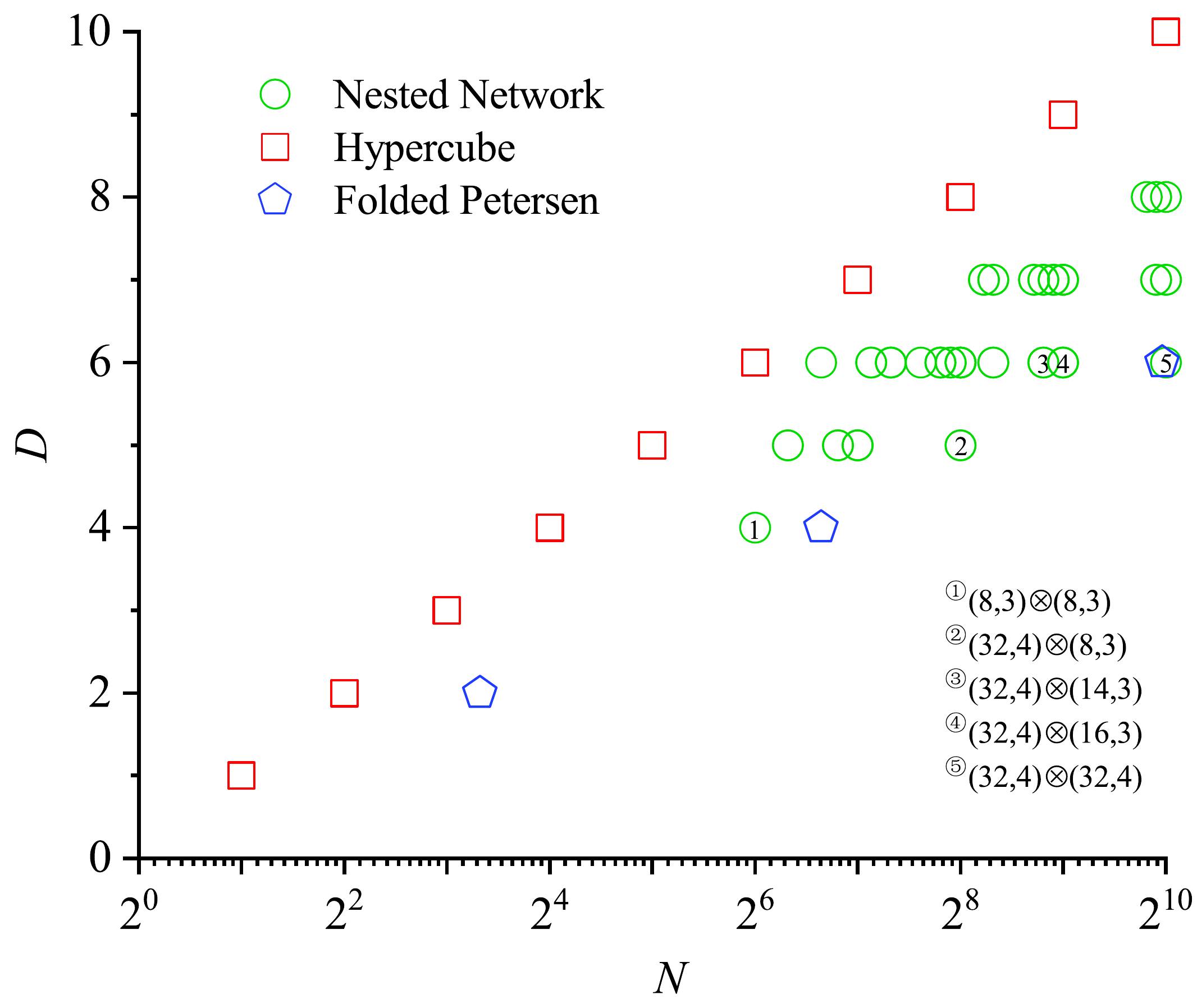}
	\caption{The diameters of the nested networks with diverse base graphs.}
	\label{multiscale}
	\end{center}
\end{figure}

For small networks, the nested network can also provide diverse choices (Fig \ref{multiscale}), the folded Petersen network generating discrete numbers of nodes $N=10,100,1000$, perform better than $(8,3)\otimes (8,3)$, $(32,4)\otimes (8,3)$, and $(32,4)\otimes(32,4)$. But our nested networks have the most diversities in network sizes and perform much better than hypercube.
\subsection{Load-balanced Routing Algorithms}
In most networks, there are dedicated processing unit for computing and messaging. While in others, one processor may need to do both, and computing and messaging can mutually impact each other. To balance such loads, we propose a load-balanced algorithm while keeping the most efficient messaging. The routing algorithm involves in two stages: (1) enumerating all the shortest paths between any two nodes and (2) choosing the path among all shortest paths according to the solutions of programming.
\par
\begin{figure}[htbp!]
\begin{center}
	\includegraphics[width=7.5cm]{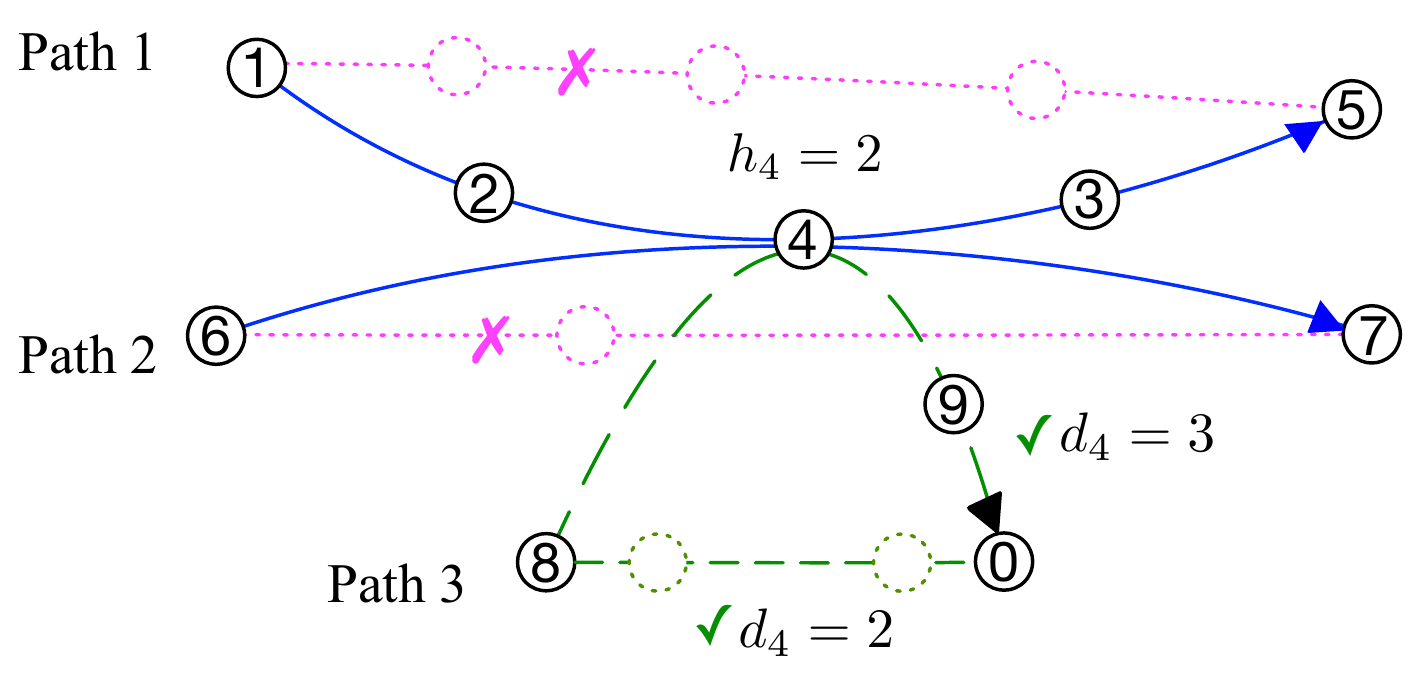}
	\caption{The definitions of variables $d_n$ and $h_n$. The dotted lines with cross sign mean that the source-destination pair only have one shortest path, and the dashed lines mean that there exist multi-path between a node pair.}
	\label{hk}
	\end{center}
\end{figure}
\begin{figure}[htbp!]
\begin{center}
	\includegraphics[width=7cm]{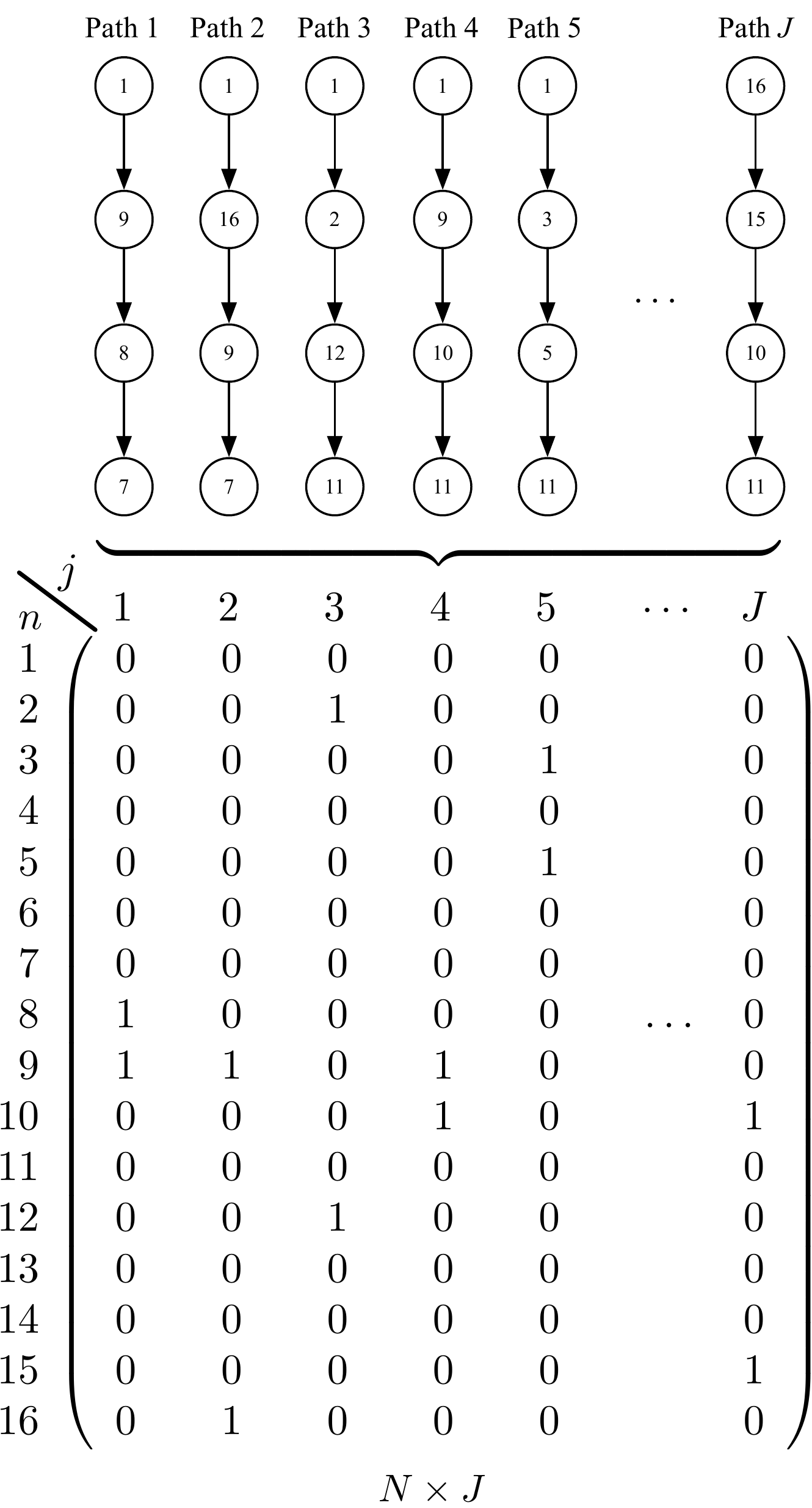}
	\caption{The definition of  matrix $P$.}
	\label{pij}
	\end{center}
\end{figure}

  
Variable $d_n$ (Fig. \ref{hk}) measures the load of node $n$, in the total number of times when the paths get through node $n$. For example, $h_4=2$ in Fig. \ref{hk}, because there exists one shortest path (dotted lines) for node pairs 1--5, 6--7, where node 4 resides. But the shortest paths, between node pair 8-0 (dashed lines), have two candidates and the node 4 is involved in one path, but not in the other one. Thus, $d_4=3$, if the upper path has been selected. For balancing the load for each node, the second stage can be written in mixed-integer nonlinear programming (MINLP), 
\begin{equation}
\label{obj}
\begin{aligned}
\min \quad{}&\sum_{n=1}^{N}\left(d_{n}-\frac{\sum d_{n}}{N}\right)^{2}\\
\text{Subject~to:}\quad &\forall s_{j} \in\{0,1\}, \forall P_{nj} \in\{0,1\},\forall \Omega_{mj} \in\{0,1\}
 \\  &\sum_{m=1}^{M}\Omega_{mj}s_j=1
  \\ & d_{n}=h_{n}+\sum_{j=1}^{J} P_{nj} s_{j}.
\end{aligned}
\end{equation}
where, $N$ is vertices and $P$ is the matrix indicated whether node $n$ resides on path $j$, but the start and end are not included. As shown in Fig. \ref{pij}, there are two paths, $1\rightarrow 9 \rightarrow 8 \rightarrow 7$ and $1\rightarrow 16 \rightarrow 8 \rightarrow 7$ for node pair 1--7. Therefore, corresponding positions of the matrix $P$ is 1. 
\par
\begin{figure}[htbp!]
\begin{center}
	\includegraphics[width=7cm]{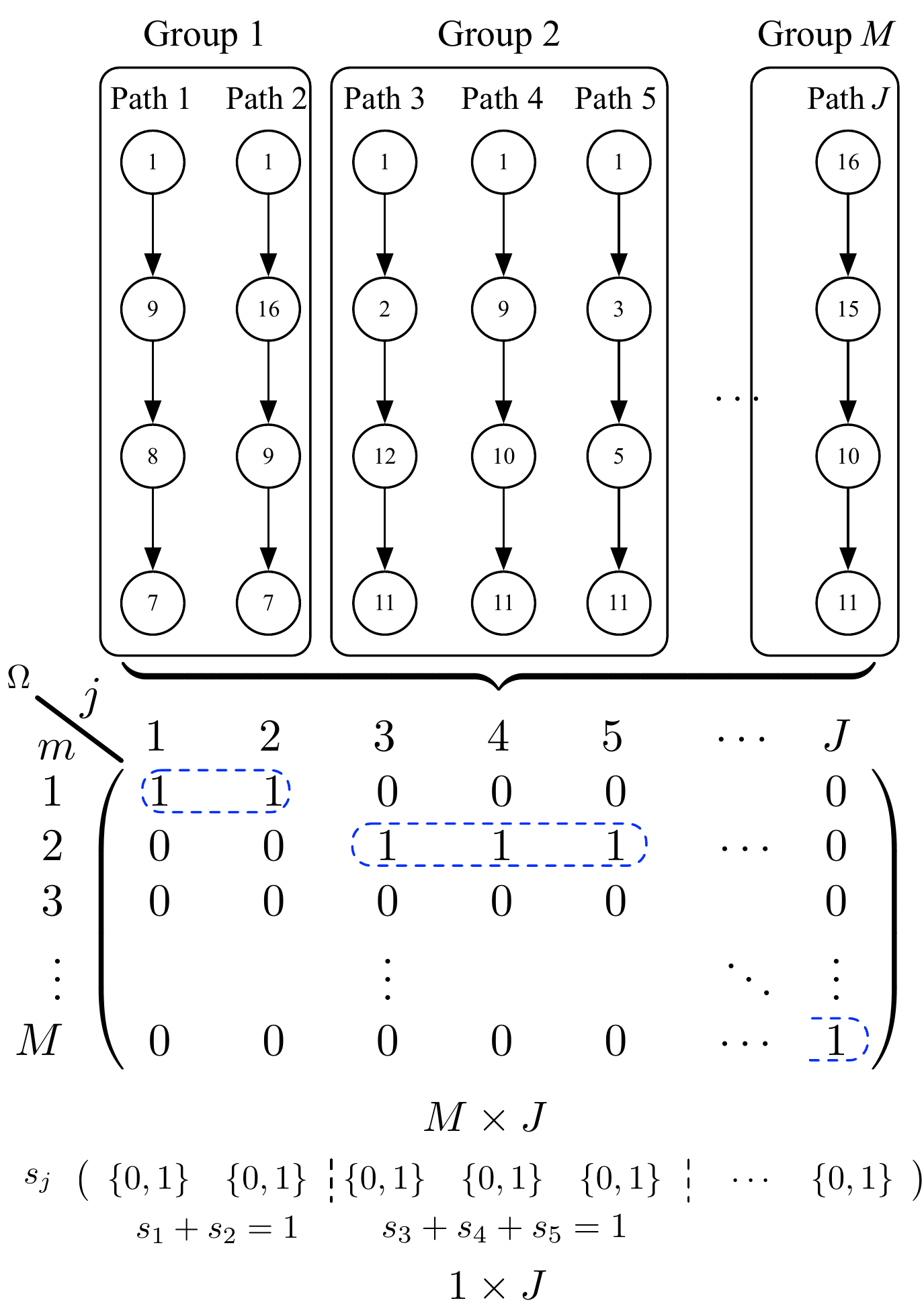}
	\caption{The legend of matrix $\Omega$ and variable $s_j$ used in the objective function (\ref{obj}).}
	\label{gij}
	\end{center}
\end{figure}

The matrix $\Omega$ is the group matrix, which puts the same source-destination pairs into one group, as shown in Fig. \ref{gij}. For example, the source-destination pair of paths 1 and 2 is $1\rightarrow 7$, so $\Omega_{11}=\Omega_{12}=1$ when we label these two paths as Group 1. In addition to the group matrix, the constraint of $\sum s_j=1$ means that there is only one path for one source-destination pair. The objective function is nonlinear, and it can be solved by using commercial or open-source packages including LINGO \cite{Neumaier2005}, IBM ILOG Cplex \cite{bliek1u2014solving}, MOSEK , \textit{etc.}, while heuristic methods such as genetic algorithm and simulated annealing are other solutions.
\par
\begin{proposition}
For product networks, if $\sum_{n=1}^{N}\left(d_{n}-\sum d_{n}/{N}\right)^{2}=0~(d_1=d_2=d_3\cdots=d_N)$ for $G_1$, $G_2$ is satisfied, for $G_1\otimes G_2$, this condition is also satisfied.  
\end{proposition}
We can prove this proposition from different type of source-destination pairs. Now we define $u \in G_1$ and $v \in G_2$, there existed a path from $\langle u_1,v_1 \rangle \rightarrow \langle u_2,v_2 \rangle$, 
\begin{enumerate}
\item $u_1=u_2$, the path is located in $G_2$ and $d_{\langle u_1,v_1\rangle}=d_{\langle u_1,v_2\rangle}=\cdots$
\item $v_1=v_2$, the path is located in $G_1$, and $d_{\langle u_1,v_1\rangle}=d_{\langle u_2,v_1\rangle}=\cdots$
\item $u_1\neq u_2$ \text{and} $v_1\neq v_2$, the path can be expressed as two paths in the $G_1$ and $G_2$ independently. 
\begin{equation}
\begin{aligned}
&\langle u_1,v_1 \rangle \rightarrow \langle u_2,v_2 \rangle\\
\Rightarrow &\langle u_1,v_1 \rangle \rightarrow \langle u_2,v_1 \rangle \rightarrow \langle u_2,v_2 \rangle\\
\text{or} \Rightarrow & \langle u_1,v_1 \rangle \rightarrow \langle u_1,v_2 \rangle \rightarrow \langle u_2,v_2 \rangle.
\end{aligned}
\end{equation}
Here, the routes of $\langle u_1,v_1 \rangle \rightarrow \langle u_2,v_1\rangle$ and $\langle u_2,v_1 \rangle \rightarrow \langle u_2,v_2 \rangle$ are proved above. 
\end{enumerate}
\par
Additionally, $\sum\left(d_{n}-\sum d_{n}/{N}\right)^{2}\neq 0$ and get the suboptimal solution, but the load of $G_1\otimes G_2$ is still near balanced. For large $N$, the size of matrix increases too rapidly to solve directly. Therefore, this feature allows load-balanced routing of larger networks from those of the small networks.
\section{Benchmarks of Routing and Networks}
Benchmarks were performed on a real cluster of 32-nodes and simulations with SimGrid \cite{Casanova2014}. For small networks, a Beowulf cluster "Taishan" \cite{Deng2019} constructed with various low-budget parts to recalibrate our results. For larger networks, the simulation provides flexible and yet reliable indication of the performance of our routing.
\subsection{Graph500 Benchmark on Taishan}
Experiments on Taishan with hardware designed for soft routers, which is a functional prototype platform for interconnection network research, the cluster contains 32 nodes with hardware configuration, (1) Processor: Intel Celeron 1037U; (2) Memory: 1$\times$8 GB General DDR3 SODIMM (1600 MHz, 1.35V); (3) Internal Storage: 128 GB General SSD; (4) Network Storage: NFS via 48-port Gigabit Ethernet switch; (5) Ethernet: Intel 82583V Gigabit Ethernet controllers (8 ports); (6) Operation System: CentOS Linux 6.7 (kernel 2.6.32); (7) Compiler: GCC 4.4.7; (8) MPI Environment: MPICH 3.2.As evident, Taishan's configuration is designed for testing network switching, instead of high-density computing. 
\begin{figure}[htbp!]
\centering
\includegraphics[width=8.5cm]{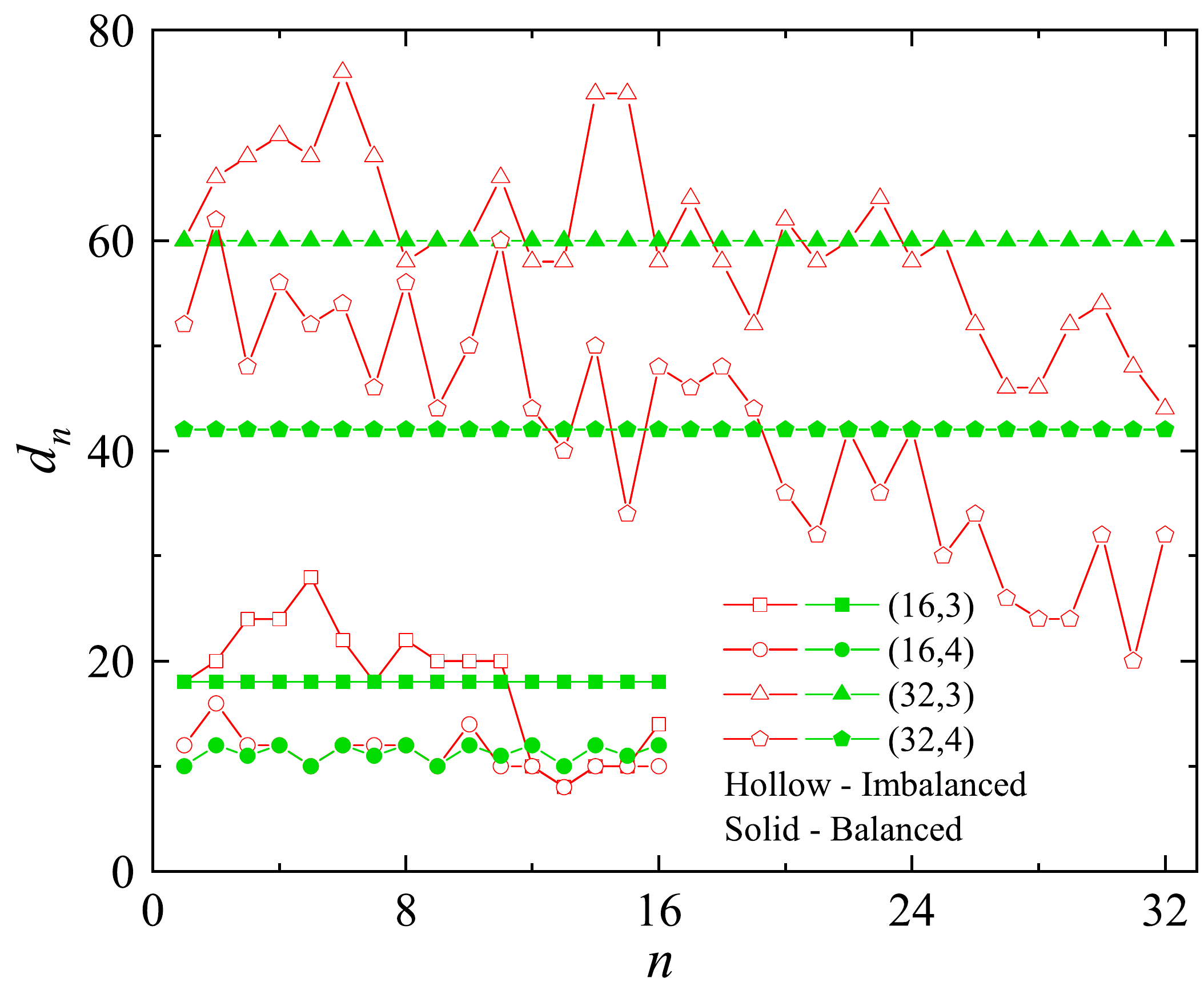}
\caption{Forwarding loads of for topologies $(16,3),(16,4),(32,3)$ and $(32,4)$ as a function of node order $n$.}\label{load-small}
\end{figure}
\par
Fig. \ref{load-small} demonstrates the differences of loading of each topology between imbalanced routing and balanced routing,  the load for each node fluctuates leading to congestion and reduction of the efficiency of the network. Except for the topology of $(16,4)$, the objective function of load-balanced routing can converge to zero, that means numbers of paths through every node is balanced, to prevent congestion on a particular node. For the topology of $(16,4)$ as shown in Fig. \ref{deng} (b), the range of load is located in the range of $(10,12)$, which improves original imbalanced routing, although the objective function is not minimized.
\par
The new routing strategy optimizes the throughput for global communication, demonstrated by Graph 500, a benchmark baseline of multiple breadth-first search (BFS) and single-source shortest path (SSSP). As shown in Tab. \ref{data}, for any topology, the balanced routing always outperforms the imbalanced ones. The gains become more obvious as we expand the network size.
\par
\begin{table*}[]
\centering
\caption{Benchmark results from an actual cluster Taishan and {SimGrid} simulations.}
\label{data}
\begin{tabular}{@{}llrrcc@{}}
\toprule
\multicolumn{2}{c}{{ \diagbox{Benchmark}{Topology}}} & \multicolumn{1}{c}{$(16,3)$}               & \multicolumn{1}{c}{$(16,4)$}               & $(32,3)$                                                    & $(32,4)$                                                    \\ \midrule
Graph500 BFS               & Imbalanced Routing      & {\color[HTML]{FE0000} $72.42\pm 0.02$}     & {\color[HTML]{FE0000} $101.72\pm 0.22$}    & \multicolumn{1}{r}{{\color[HTML]{FE0000} $113.53\pm 0.28$}} & \multicolumn{1}{r}{{\color[HTML]{FE0000} $169.45\pm 0.61$}} \\ \cmidrule(l){2-6} 
($10^6 $ MTEPS)             & Balanced Routing        & {\color[HTML]{036400} $84.63\pm 0.06$}     & {\color[HTML]{036400} $102.96\pm 0.11$}    & \multicolumn{1}{r}{{\color[HTML]{036400} $124.00\pm 0.36$}} & \multicolumn{1}{r}{{\color[HTML]{036400} $191.45\pm 0.22$}} \\ \midrule
Graph500 SSSP              & Imbalanced Routing      & {\color[HTML]{FE0000} $37.27\pm 0.27$}     & {\color[HTML]{FE0000} $48.45\pm 0.41$}     & \multicolumn{1}{r}{{\color[HTML]{FE0000} $59.70\pm 0.40$}}  & \multicolumn{1}{r}{{\color[HTML]{FE0000} $77.15\pm 0.86$}}  \\ \cmidrule(l){2-6} 
($10^6$ MTEPS)             & Balanced Routing        & {\color[HTML]{036400} $41.14\pm 0.34$}     & {\color[HTML]{036400} $48.58\pm 0.42$}     & \multicolumn{1}{r}{{\color[HTML]{036400} $64.67\pm 0.43$}}  & \multicolumn{1}{r}{{\color[HTML]{036400} $85.22\pm 0.78$}}  \\ \midrule \midrule
\multicolumn{2}{c}{{ \diagbox{Benchmark}{Topology}}} & \multicolumn{1}{c}{$(16,3)\otimes (16,3)$} & \multicolumn{1}{c}{$(16,4)\otimes (16,4)$} & \multicolumn{1}{l}{}                                        & \multicolumn{1}{l}{}                                        \\ \midrule
Bandwidth Efficiency       & Imbalanced Routing      & {\color[HTML]{FE0000} 5.08}                & {\color[HTML]{FE0000} 6.35}                &                                                             &                                                             \\ \cmidrule(l){2-6} 
($10^3$ MB/s)              & Balanced Routing        & {\color[HTML]{036400} 6.10}                & {\color[HTML]{036400} 7.22}                &                                                             &                                                             \\ \midrule
NPB FT-Class A             & Imbalanced Routing      & {\color[HTML]{FE0000} 20.59}               & {\color[HTML]{FE0000} 24.84}               &                                                             &                                                             \\ \cmidrule(l){2-6} 
($10^3$ Mops/s)            & Balanced Routing        & {\color[HTML]{036400} 21.95}               & {\color[HTML]{036400} 26.90}               &                                                             &                                                             \\ \midrule
NPB FT-Class C             & Imbalanced Routing      & {\color[HTML]{FE0000} 41.3}                & {\color[HTML]{FE0000} 53.45}               &                                                             &                                                             \\ \cmidrule(l){2-6} 
($10^3$ Mops/s)            & Balanced Routing        & {\color[HTML]{036400} 61.4}                & {\color[HTML]{036400} 79.24}               &                                                             &                                                             \\ \bottomrule
\end{tabular}
\end{table*}
\subsection{Benchmarks of Large Networks by Simulation}
The evaluation of larger-scale topologies generated from product networks is carried out with SimGrid, a package that provides the accurate, versatile, and scalable simulation for distributed computing or cloud platform, especially with SMPI, help simulate MPI applications with little modification \cite{Casanova2014}. The parameters selected for our simulations are 8 Gflops core processing speed, full-duplex gigabit link and 30 $\mu$s latency per link, and they are so chosen as to model the "Taishan" cluster in actual but much smaller setting. Two static routing tables, one with Floyd algorithm without balanced routing, and the other one is balanced routing generating from small graphs, we selected.
\begin{figure}[htbp!]
\centering
\includegraphics[width=9cm]{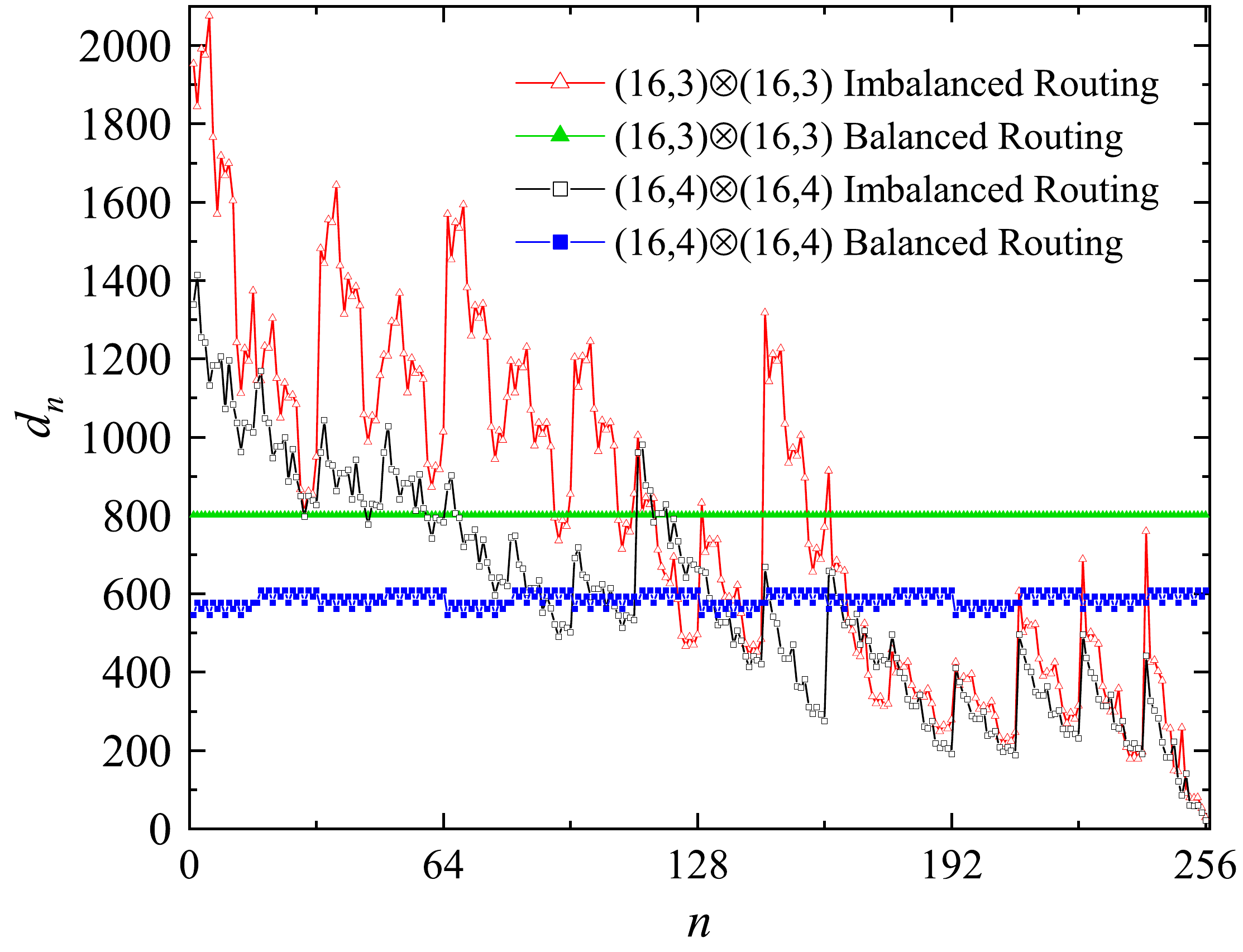}
\caption{Messaging loads of $(16,3)\otimes (16,3)$ and $(16,4)\otimes (16,4)$.}\label{pro-load}
\end{figure}
\par
Fig. \ref{pro-load} shows the loads for two two routing schemes on two nested networks of $(16,3)\otimes (16,3)$ and $(16,4)\otimes (16,4)$. As represented by green line, each node has the same load, resulting from the load-balanced routing of $(16,3)$ (Fig. \ref{load-small}), the Floyd algorithm leads to extensive load fluctuations, that causing congestion. As shown in blue line, the balanced routing fluctuates slightly due to the suboptimal solution for load-balanced routing of $(16,4)$ as in Fig. \ref{load-small}. The balanced routing performs much smoothly than the Floyd imbalanced routing.
\par
\begin{figure}[htbp!]
\centering
\includegraphics[width=8cm]{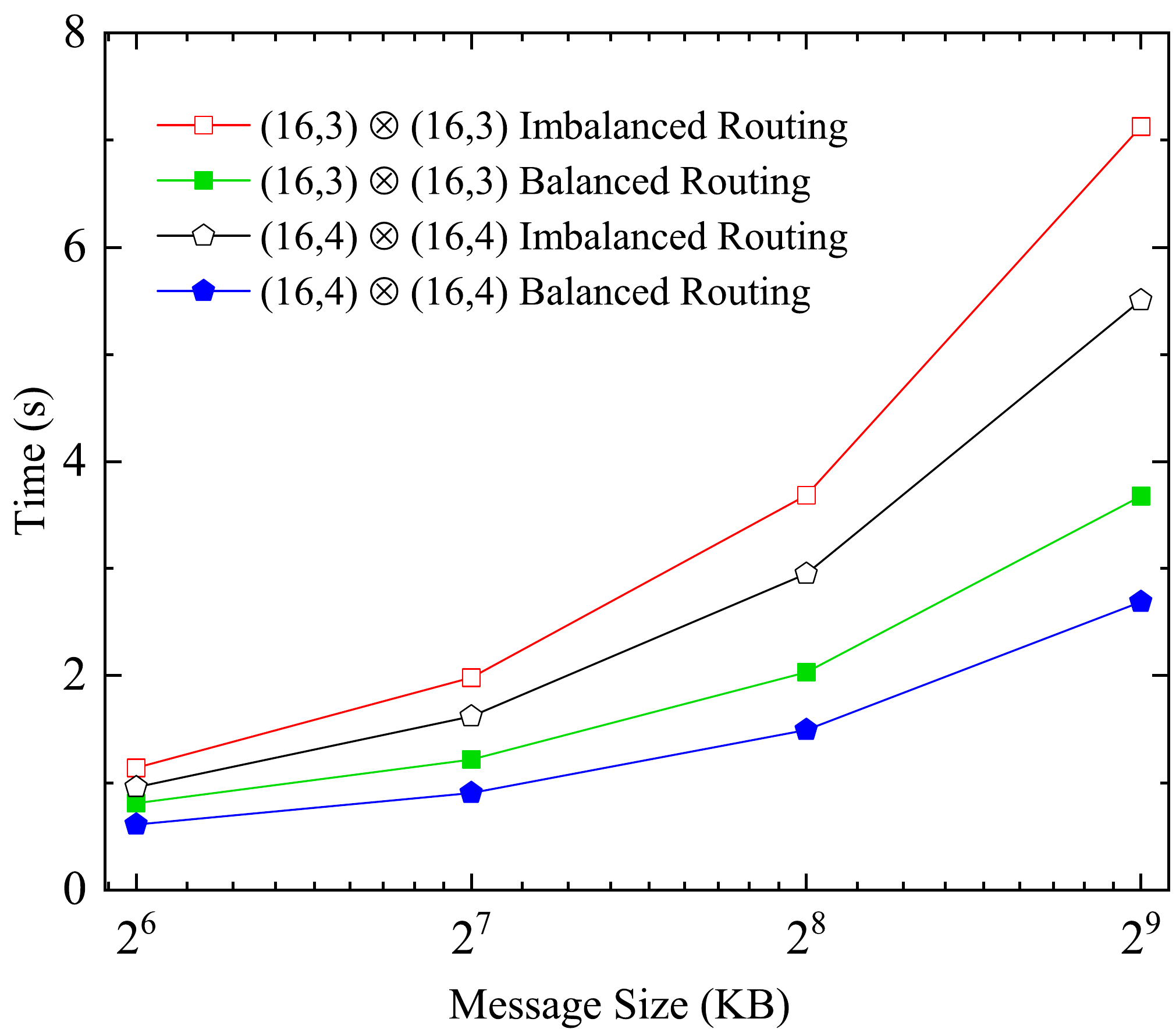}
\caption{Benchmark results for MPI All-to-all.}\label{mpiall2all}
\end{figure}
With balanced routing, we tested MPI All-to-all, effective bandwidth, and NPB FT, all of them depend heavily on global communication, to survey the effect of balanced routing algorithm. Additionally, almost all network simulators including SimGrid and NS \cite{Falla} only support a single core, with the 128 GB RAM for computing nodes on SeaWulf, that limit the scale of topology and complexity of applications, especially for high-density computing. Hence, the nested networks of $(16,4)\otimes(16,4)$ and $(16,3)\otimes(16,3)$ are be used to evaluate the performance of the balanced routing, while unit message sizes for MPI All-to-All are not over 512 KB, maximum message size for effective bandwidth benchmark is 1MB, and the NPB FT (3D-FFT benchmark) use two types of Class A, whose size is $256\times 256\times 128$, and Class C, whose size is $512\times 512\times 512$. 
\par
Fig. \ref{mpiall2all} presents the results of MPI All-to-all, the messaging time decreases dramatically, to half of those with imbalanced routings, especially for small packets. Tab. \ref{data} shows the bandwidth increases for two networks of $(16,3)\otimes(16,3)$, and $(16,4)\otimes(16,4)$ with imbalanced or balanced routings. The relative gains range from 20.0\% for $(16,3)\otimes(16,3)$ to 13.8\% for $(16,4)\otimes(16,4)$. For the benchmark of NPB FT, performances increase tremendously, especially for Class C where more data are involved. All benchmarks and the forwarding loads for $(16,3)\otimes(16,3)$ and $(16,4)\otimes(16,4)$ show that high network degrees help reduce the forwarding loads and improve the network throughput. 
\par
In summary, benchmark experiments on Taishan and numerical simulations on much larger clusters with SimGrid demonstrated the superiority of the balanced routing for our networks and this routing can be generalized naturally for multi-level product networks for constructing much larger networks, that is critical to cloud systems as well as the next-generation supercomputers.
\section{Conclusions}
We propose a scheme to multiply base and easy-to-optimize networks to construct larger and scalable networks with desired properties for a variety of applications in, for example, supercomputers and large cloud systems. To help realize the potentials of these networks, we introduced a balanced routing method that enable load balance for static routing, as messaging load balance is a highly visible pitfall of the product graphs. To demonstrate the values of balanced routing, recalibration of Graph500 for our previous work on the enhanced networks and simulations for product networks, we performed a series of benchmarks and obtained expected gains of performances in all cases for the product networks coupled with our balanced routing.
\section*{Acknowledgments}
We thank the National Supercomputing Center in Jinan for support of constructing the ”Taishan” cluster. We also thank Stony Brook Research Computing and Cyberinfrastructure, and the Institute for Advanced Computational Science at Stony Brook University for access to the high-performance SeaWulf computing system, which was made possible by a \textdollar 1.4M National Science Foundation grant (\#1531492). Also, we would like to thank the following individuals for their contributions to this work at different stages, M. Guo, W. Liu, L. Pei, and W. Cun.

\ifCLASSOPTIONcaptionsoff
  \newpage
\fi



\bibliographystyle{IEEEtran}
%
\bibliography{Refs.bib}

%

\end{document}